%% file: main.tex
\documentclass[sigconf]{acmart}

\AtBeginDocument{%
  \providecommand\BibTeX{{%
    \normalfont B\kern-0.5em{\scshape i\kern-0.25em b}\kern-0.8em\TeX}}}

\usepackage{natbib}%
\usepackage{multirow}
\usepackage{tabularx}
\usepackage[normalem]{ulem}
\usepackage{hyperref}
\usepackage{graphicx}
\usepackage{tcolorbox}
\usepackage{underscore}
\usepackage{booktabs}

\begin{document}

\title{Open Source Software: An Approach to Controlling Usage and Risk in Application Ecosystems}

\input{authors}

\renewcommand{\shortauthors}{Zajdel et al.}

\newcommand{\Zajdel}[1]{\textcolor{green}{{\it [Zajdel: #1]}}}
\newcommand{\hafedh}[1]{\textcolor{blue}{{\it [Hafedh: #1]}}}
\newcommand{\Costa}[1]{\textcolor{red}{{\it [Costa: #1]}}}
\newcommand{\diego}[1]{\textcolor{red}{{\it [Diego: #1]}}}

\newcommand{\code}[1]{\texttt{#1}}
\newcommand{\tool}[1]{\textsc{GitHub-Miner}}

\input{abstract}



\maketitle

\input{sections/1-introduction}

\input{sections/2-problem-with-open-source}
\input{sections/3-using-automation-to-reduce-open-source-risk}
\input{sections/4-results}

\input{sections/5-discussion}

\input{sections/6-relatedwork}

\input{sections/7-conclusion}

\bibliographystyle{acm/ACM-Reference-Format}
\bibliography{bibliography.bib}

\end{document}

%% file: authors.tex
\author{Stan Zajdel}
\email{Stan.Zajdel@ibx.com}
\orcid{1234-5678-9012}
\affiliation{%
  \institution{Independence Health Group}
  \streetaddress{1901 Market Street}
  \city{Philadelphia}
  \state{PA}
  \country{USA}
  \postcode{19103}
}

\author{Diego Elias Costa}
\email{diego.costa@concordia.ca}
\affiliation{%
  \institution{Concordia University}
  \streetaddress{1919 Market Street}
  \city{Montréal}
  \state{Québec}
  \country{Canada}
  \postcode{19103}
}

\author{Hafedh Mili}
\email{mili.hafedh@uqam.ca}
\affiliation{%
  \institution{LATECE lab, Université du Québec à Montréal}
  \streetaddress{P.O. Box 8888, downtown station}
  \city{Montréal}
  \state{Québec}
  \country{Canada}
}

%% file: abstract.tex
\begin{abstract}
The Open Source Software movement has been growing exponentially for a number of years with no signs of slowing. Driving this growth is the wide-spread availability of libraries and frameworks that provide many functionalities. Developers are saving time and money incorporating this functionality into their applications resulting in faster more feature-rich releases. Despite the growing success and the advantages that open source software provides, there is a dark side. Due to its community construction and largely unregulated distribution, the majority of open source software contains bugs, vulnerabilities and other issues making it highly susceptible to exploits. The lack of oversight in general hinders the quality of this software resulting in a trickle down effect in the applications that use it. Additionally, developers who use open source tend to arbitrarily download the software into their build systems but rarely keep track of what they have downloaded resulting in an excessive amount of open source software in their applications and in their ecosystem. This paper discusses processes and practices that users of open source software can implement into their environments that can safely track and control the introduction and usage of open source software into their applications, and report on some preliminary results obtained in an industrial context. We conclude by discussing governance issues related to the disciplined use and reuse of open source and areas for further improvements.
\end{abstract}

%% file: sections/1-introduction.tex
\section{Introduction}
The open source software industry has experienced an explosion of growth in the last several years. It is predicted that the open source software market will continue to grow at a Compound Annual Growth Rate (CAGR) of 17\% between 2021-27 and is expected to reach USD \$60 billion by 2026 (1). Fueling this growth is the widespread adoption of open source software by businesses and organizations~\cite{OpenSour5:online}.

The last several years have ushered in many changes in the way organizations are approaching software development and delivery. To be more competitive, organizations have been increasingly adopting practices such as Agile and DevOps to speed up development and delivery of applications. The use of open source software has enabled this growth by making available freely reusable capabilities that can be easily integrated into applications thus reducing time to market. Reports estimate that between 90\% to 93\% of all businesses and organizations are using open source software in their applications and systems~\cite{TheValue30:online,tidelift:19}. 

However, the unfettered use of open source software has a major downside: organizations are increasingly relying, for their mission critical systems, on software of uncertain quality, developed under processes of uneven quality. In his seminal--and much cited--cathedral versus bazaar paper \cite{Raymond1999}, Raymond argued convincingly that several factors, characteristic of large and active open source projects, contribute to the quality of the resulting software, including the (very) large number of users, and the fact that these users are themselves--often sophisticate--developers. While these conditions held for Unix and its derivatives--the textbook case of large, complex, and successful open source software--they are actually the exception rather than the rule. Further, there is a qualitative difference between functional and performance bugs, on the one hand, and security vulnerabilities, on the other. Open source developers do not actively look for functional or performance bugs: they encounter them in the context of their testing and production. Security "bugs", on the other hand, are of a different nature: malicious intent is needed to exploit what often amounts to an implementation's permissiveness beyond the original functional specification. A recent study indicates that up to 40\% of the Node.js packages rely on code with known vulnerabilities~\cite{Zimmermann:19:Smallworld}, which can affect the security and stability of the applications dependent on this software (3).

While organizations cannot control the \textit{upstream} processes used to develop the reusable software they integrate in their own products, they can at least control \textit{downstream} which reusable components go into their products, based on \textit{documented} or \textit{suspected} security vulnerabilities. The challenge is to do so in an efficient manner, and in a way that does not break the agility of their own software development processes. This paper reports on the efforts of an organization--a sizable health insurance corporation--to address this challenge, through a combination of \textit{automation}, \textit{education}, and \textit{governance}. Our strategy rests on the following hypotheses and principles:
\begin{itemize}
    \item ($H_1$): open source software (OSS) is of uneven quality
    \item ($H_2$): there is enough information out there, both direct, and indirect, to assess the quality of available OSS
    \item ($P_1$): we should strive to minimize the disruption to the work flow of developers, but
    \item ($H_3$): adequately 'educated' developers can help reduce the disruption.
\end{itemize}
Developers have come to rely on large ecosystems of OSS libraries to deliver feature-rich applications in record times, by focusing on domain functionality and reusing the rest. Principle $P_1$ says that we should not break this agility in the process of improving OSS quality. Our preliminary results show that this is possible (see Section \ref{sec:results}).

This paper focuses on the \textit{automation} component of our strategy, with \textit{education} and \textit{governance} briefly discussed in Section \ref{sec:discussion}. Per principle $P_1$, we let developers use whichever OSS library gets the job done, and then flag problematic ones during the DevOps pipeline--we then talk about Dev\underline{Sec}Ops pipelines (see Section \ref{subsec:from-devops-to-devsecops}). This is done by parsing build scripts to identify first order dependencies (see Section \ref{subsec:dependency-reference-service}), and querying a database that documents OSS libraries that are known to the organization. This database combines information gleaned from the (US) National Vulnerability Database (NVD\footnote{\url{https://nvd.nist.gov}}), and from other sources, including the results of our own investigations. Libraries entered into our database are in one of three states: 1) \textit{allowed}, if they are know to be of good quality, with no or few security vulnerabilities, 2) \textit{disallowed}, if they have documented security vulnerabilities, and 2) \textit{under evaluation}, with self-explanatory meaning. The first time a library is encountered in a build, it is put \textit{under evaluation}, and the owner of the build is notified (see Section \ref{subsec:new-software-vetting-process}). If a build includes a \textit{disallowed} library, it is given a reprieve, the first time it is submitted, with the owners of the build notified and urged to remove or replace the faulty library within a period of time, after which the build will fail (see Section \ref{subsec:new-software-vetting-process}), unless the developers present a very good argument of why it should be allowed.  

The next section provides an overview of the problems raised by the extensive availability and use of OSS libraries in mission critical applications (Section \ref{sec:problem-with-open-source-software}), and introduces the company context (Section \ref{subsec:company-context}); to the extent that a single organization cannot control the 'supply chain' of open source software (Section \ref{subsec:open-source-problem-supply-chain}). Our solution focuses on the developers' usage of OSS libraries (Section \ref{subsec:open-source-problem-developer-usage}). 
Section \ref{sec:using-automation} presents our solution that relies on automation within the context of a DevSecOps pipeline. Our DevSecOps pipeline, which implements part of the vision described in Section \ref{sec:using-automation}), has been in operation for X months and has shown great promise in reducing the security exposure of our applications, with very little effect on the productivity of developers; the preliminary results are presented in section \ref{sec:results}. We discuss future steps towards completing the DevSecOps pipeline (\textit{automation}), but also towards developer education and enhanced governance, in Section \ref{sec:discussion}. Related work is discussed in Section \ref{sec:related-work}; we conclude in Section \ref{sec:conclusion}

%% file: sections/2-problem-with-open-source.tex
\section{The Problems with Open Source} 
\label{sec:problem-with-open-source-software}

In this section, we describe two groups of problems related to open source software. 
We start by describing the problems at the open source supply chain level (Section~\ref{subsec:open-source-problem-supply-chain}) and then describe problems with developer usage of that software (Section~\ref{subsec:open-source-problem-developer-usage}). Each are resulting in excessive risk to the security, quality and maintainability of applications that use that software. 

\subsection{Problems at the Open Source Supply Chain Level}
\label{subsec:open-source-problem-supply-chain}

The open source software community is generally an unregulated industry. 
The majority of projects are maintained by either a single developer or a small team of volunteer developers, motivated by intrinsic and internalized motivators~\cite{Gerosa:21:Motivation}. 
How often they actually have the time and resources to properly test and maintain their code is virtually unknown, and not subject to any formal process.
Many open source projects are not properly maintained~\cite{Kalliamvakou:14:PerilsGithub,Coelho:20:Maintained} and lack formal testing~\cite{Abdalkareem:17:Trivial}, which puts the responsibility of ensuring code quality on their users. 
The lack of documented proof that the software has been rigorously tested for both security and quality at the community level is contributing to many problems - some of the major ones are described below:

\begin{enumerate}
   \item \textbf{Bugs and Vulnerabilities}: At the top of the list are the bugs and vulnerabilities that are prevalent in any type of software. 
   The general lack or tailored security testing in open source development exacerbates this problem. 
   While it's been argued that the sheer number of open source contributors are actually limiting the number of bugs and vulnerabilities the flip side of that argument is the fact that there are almost an equal number of bad actors who are also downloading and analyzing that code to discover ways to exploit and attack the software~\cite{Alfadel:21:Python,Decan:18:NPM}. 
   The recent log4shell issue, among many others, is a prime example of this situation and it was fortunate that the issue was discovered and prudently reported by a researcher thus preventing a world-wide disaster~\cite{Log4Shel70:online}.

   \item \textbf{Intentional Sabotage}: Another problem that is prevalent in open source software is the intentional sabotaging of the software by contributors~\cite{Forsgreen:20:Octoverse}. One recent example of this was the intentional sabotaging of two very popular libraries, Color and Faker, by the maintainer who was disgruntled at how profitable corporations were using open source software. The maintainer deliberately added an endless loop into the Colors library and it was subsequently downloaded roughly 23 million times~\cite{Devcorru8:online}. 
   Another more recent example of intentional sabotage happened on a popular library called node-ipc~\cite{nodeipcN83:online}. 
   The maintainer, in protest of the Russian invasion of the Ukraine, maliciously added code to replace files with a heart emoji and a peace-not-war module. The malicious code was designed to specifically target users with IP addresses located in Russia or Belarus. An act like this could easily be targeted everywhere. This highlights, again, the general lack of oversight and quality testing that would have most likely caught these issue prior to being uploaded into the public repository.

   \item \textbf{Repository Poisoning:} Repository poisoning is another issue that has occurred in the repositories that house open source code such as Maven, NPM and PyPI~\cite{Zimmermann:19:Smallworld,Vu:20:TyposquattingPython,Forsgreen:20:Octoverse}. 
   Bad actors are known to create libraries that mimic valid libraries and intentionally add vulnerable code in them~\cite{Zimmermann:19:Smallworld}. The libraries are then renamed slightly different to closely match the legitimate versions and uploaded to the repository. Developers arbitrarily download those poisoned libraries without paying much attention to the name which results in the unnoticed introduction of an exploitable library into their code base.

   \item \textbf{Inadequate Documentation:} Additional issues include an overall lack of adequate documentation for much of the open source software. According to a recent GitHub survey, 93\% of the respondents noted that incomplete or outdated documentation is a pervasive problem leading to misunderstandings on how to use that software ultimately leading to application bugs and instability~\cite{OpenSour92:online}.

   \item \textbf{Software Bloat:} The collaborative nature of open source software, while a good thing, also has it's drawbacks - one of them being that it can lead to bloat. For many projects there is no central authority to monitor which features are being added to the software. This can lead to a number of issues such as bloat or a drift from the intended requirement for the software. This can lead to an increased attack surface, unintentional bugs or an overall lack of quality that can negatively affect the applications using this software~\cite{Ponta:21:Bloated}.
\end{enumerate}

\subsection{Problems with Developer Usage of OSS}
\label{subsec:open-source-problem-developer-usage}

It has been our experience that developers who use OSS tend to arbitrarily download the software into their build systems but rarely keep track of what they have downloaded, and even when they do, they hardly keep track of which versions they use \cite{Kula2018}. This happens during normal feature development cycles where developers are exploring the solutions space. Upon settling on a specific solution, they seldom go back and remove the other unused software that they've added. The result is unnecessary software being left in an application which can potentially increase the risk of attack. This problem is exacerbated when other developers are doing the same thing in their projects ultimately leading to a lot of excess software in the application ecosystem. 

Developers also tend to favor the open source libraries that they are comfortable in using or due to the popularity level of that specific software \cite{Kula2018,Mujahid:21:Centrality}. This leads to many different libraries that provide similar capabilities that clutter the application ecosystem. This also adds risk to the overall application environment and cost of maintainability in that developers are required to support the use of unfamiliar libraries. 

The problems as described above are resulting in organizations assuming tremendous security and stability risks as well as an increase in overall maintenance costs. It is unrealistic to expect organizations to abandon open source. On the contrary, the use is going to continue to grow and correspondingly the number of issues are going to increase as well therefore it falls upon the users of open source to find solutions to reduce that risk. Automation is the key.

\input{sections/2.3-company-context}

%% file: sections/2.3-company-context.tex
\subsection{Company Context}
\label{subsec:company-context}

The organization that is referenced in this paper is a large national insurance firm with a diverse application portfolio. The majority of the application portfolio has been written in house by a development staff of 400 software developers. The organization has been leveraging open source software for a number of years and have recently adopted agile and DevSecOps practices. The adoption of DevSecOps has lead to the widespread exposure of vulnerabilities and other issues within the application ecosystem, hence the need for a solution to control open source usage.

%% file: sections/3-using-automation-to-reduce-open-source-risk.tex
\section{Using Automation to Reduce Open Source Risk}
\label{sec:using-automation}
\subsection{From DevOps to DevSecOps}
\label{subsec:from-devops-to-devsecops}

Short of controlling the uncontrollable--the way OSS is built and its quality assured (see Sec. \ref{subsec:open-source-problem-supply-chain})--we can control it use (see Sec. \ref{subsec:open-source-problem-developer-usage}). Recall, from the introduction, that our approach rests on the following hypotheses and principles:
\begin{itemize}
    \item ($H_1$): open source software (OSS) is of uneven quality
    \item ($H_2$): there is enough information out there, both direct, and indirect, to assess the quality of available OSS
    \item ($P_1$): we should strive to minimize the disruption to the work flow of developers, but
    \item ($H_3$): adequately 'educated' developers can help reduce the disruption.
\end{itemize}
Per principle $P_1$, it is critical that we do not break the agility of developers while they explore the solution space for their development problems. Thus, we should not exert our control of OSS usage \textit{downstream} from the highly creative phase of coding, but \textit{upstream} from deployment. \textit{DevOps} is a set of practices and tools that aims at automating the continuous delivery of new software updates while ensuring their correctness and reliability \cite{Leite2019}. 
Automated build and deployment \textit{pipelines} are an integral part of a DevOps practice. \textit{DevSecOps}, as a practice, is an extension of DevOps that enables development (Dev), operations (Ops) and \textit{security} (Sec) teams, to work jointly to ensure the continuous delivery of software updates that are devoid of security vulnerabilities. Thus, we argue that the proper place to control OSS usage is at the CI pipeline.

A typical continuous integration (CI) DevSecOps pipeline in illustrated in Figure \ref{figure:pipeline-automation} and involves the following steps: 

\begin{figure} 
    \centering
    \includegraphics[width=\linewidth,trim=0.2cm 22cm 4cm .2cm, clip]{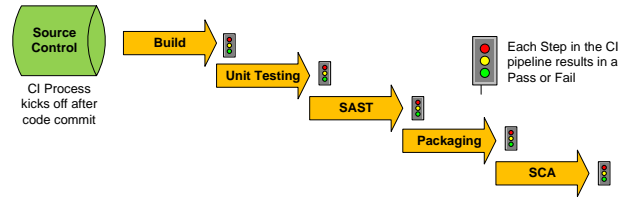}
    \caption{DevOps automation pipeline}
    \label{figure:pipeline-automation}
\end{figure}

\begin{enumerate}
   \item Code is committed to a repository
   \item CI pipeline automatically detects the code commit and starts a build
   \item After the Build completes, unit tests are run
   \item Static Application Security Testing (SAST) is performed
   \item The code is packaged
   \item Source Composition Analysis (SCA) is performed on the package
   \item The package is deployed to an internal repository
\end{enumerate}

\textit{Static Application Security Testing (SAST)} consists of checking the code base against a set of coding rules that embody best security practices--or conversely, 'security smells'--as defined by the tool. Popular tools on the market include SonarQube, Veracode, Checkmarx and Synopsis. 

\textit{Source Composition Analysis (SCA)} provides visibility into the open source components and libraries that are incorporated into application build systems. Source composition analysis helps to identify licensing issues and vulnerabilities by cross-referencing the information in such sources as the National Vulnerability Database (NVD) and others. Popular tools include Nexus IQ, Snyk and Synopsis.

\begin{figure}
    \centering
    \includegraphics[width=\linewidth,trim=0.2cm 22cm 2cm .2cm, clip]{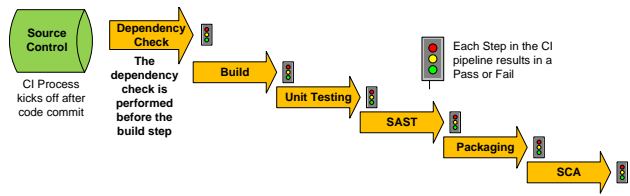}
    \caption{Identifying dependencies in a build in the DevOps pipeline}
    \label{figure:pipeline-we-service-automation}
\end{figure}

While such CI pipelines have been shown to be fairly effective at exposing bugs, vulnerabilities and other issues, they suffer from one major problem: the source composition analysis (SCA) is performed way too late in the process, with two major consequences:
\begin{itemize}
    \item The process is wasteful. To the extent that the inclusion of a given library, or a specific version thereof, with a critical security vulnerability is usually a show-stopper, it seems wasteful to work on passing the earlier gates (unit tests, SAST, packaging), to have the work thrown away
    \item Once an application is built, functional, and packaged, it may be tempting to overlook security concerns and deploy anyway, "until we can find the time to fix it".
\end{itemize}
Accordingly, we chose to modify the typical DevSecOps pipeline by moving upstream the control of the usage of OSS: in fact, OSS usage control is now the \textit{first step} in the DevSecOps pipeline, as illustrated in Figure \ref{figure:pipeline-we-service-automation}.

\begin{figure} [H]
    \centering
    \includegraphics[width=\linewidth,trim=0.2cm 20cm 6cm 0.8cm, clip]{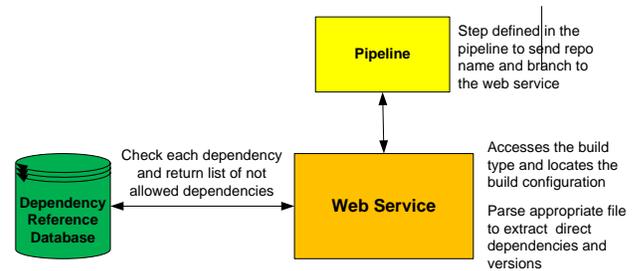}
    \caption{Process of recording dependencies in the Dependency Reference Database (RDR) during a Build}
    \label{figure:identify-dependencies}
\end{figure}

Specifically, we envision a (RESTful) \textit{service} that is triggered by source control commits and that performs the following (see Figure \ref{figure:identify-dependencies}):
\begin{enumerate}
    \item It analyzes the application being built to identify its dependencies. It first identifies the build type, to then locate the build configuration files from which it can read/retrieve the dependencies.
    \item It checks the uncovered dependencies against a \textit{Dependency Reference Database} that contains information about known libraries, including known vulnerabilities, their severity level, and eventual remediation strategies.
    \item Based on its findings, it can apply organization-specific business rules to take remedial actions.
\end{enumerate}

In the remainder of this section, we will elaborate on, 1) the contents of \textit{Dependency Reference Database} (DRD) (Section \ref{subsec:dependency-reference-database}), and 2) a web service used primarily to populate the \textit{Dependency Reference Database} (DRD), called the \textit{Dependency Reference Service} (DRS) (Section \ref{subsec:dependency-reference-service}).
\subsection{The Dependency Reference Database (DRD)}
\label{subsec:dependency-reference-database}
\subsubsection{SBOM - A Software Bill of Materials}
\label{subsubsec:SBOM}
A software bill of materials (SBOM) is a list of components in a piece of software. Software vendors often create products by assembling open source and commercial software components. The SBOM describes the components in a product. It is analogous to a list of ingredients on food packaging  where you might consult a label to avoid foods that may cause an allergic reaction. SBOMs can help organizations or persons to avoid consumption of software that could harm them [5].

Most source composition analysis (SCA) tools, which scan libraries for known vulnerabilities, keep track of an SBOM for the applications that are being analyzed which usually includes both direct and indirect, or transitive, dependencies. A transitive dependency is a dependency that is introduced by the direct dependencies that a developer physically includes in their software build configurations. 

\begin{figure}
    \centering
    \includegraphics[width=\columnwidth]{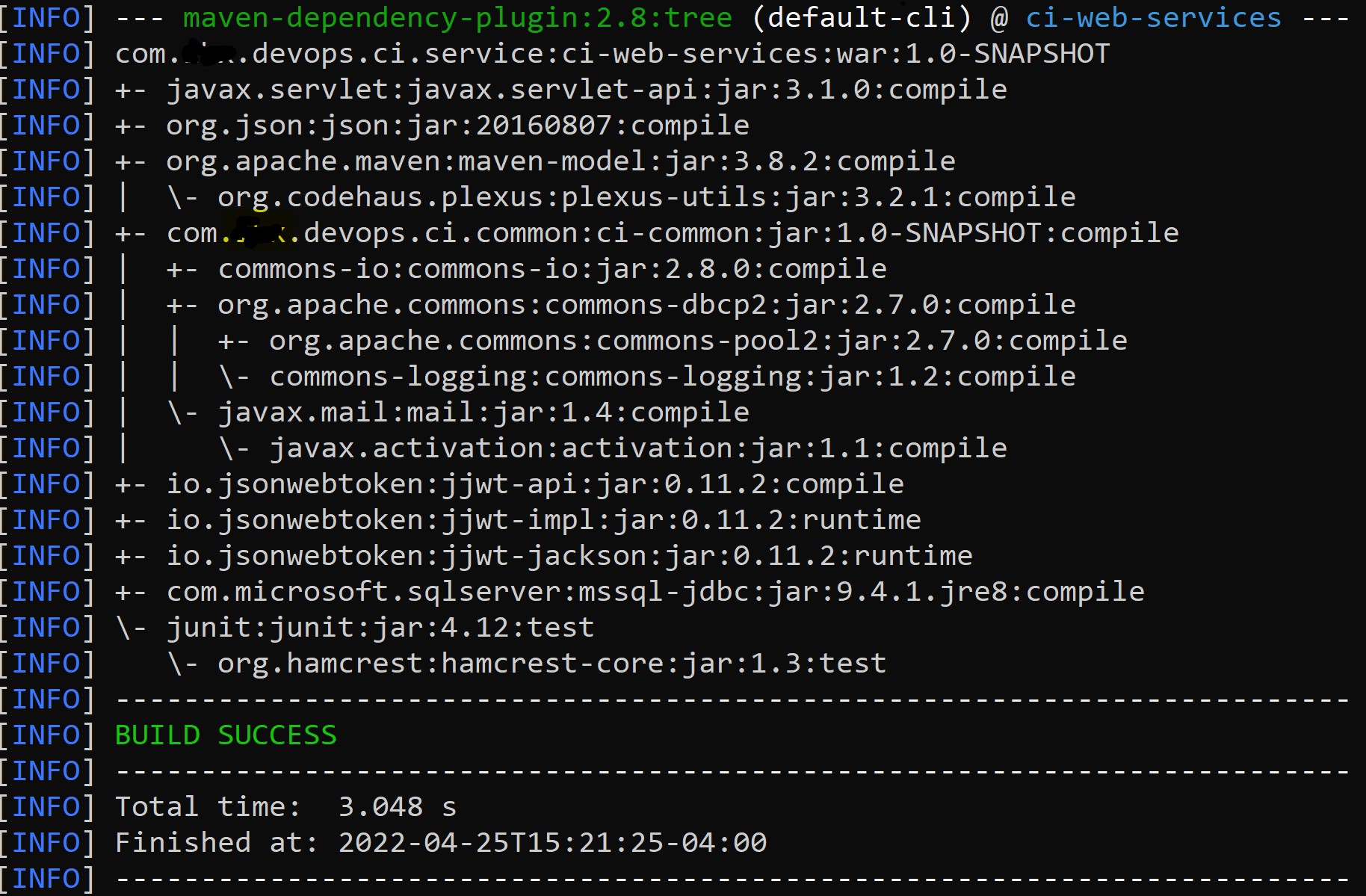}
    \caption{Maven Dependency Tree}
    \label{figure:dependency-tree}
\end{figure}

In this approach however, a different kind of SBOM is used - only direct dependencies are tracked. The reason why this solution is only concerned about direct dependencies is because a direct dependency and it's version are controlled by the developer whereas a transitive dependency depends on library maintainers' decisions. 
Hence, the SBOM, as defined by this solution, is designed to track solely direct dependencies and the versions of those dependencies that are in use. Figure~\ref{figure:dependency-tree} shows a typical Maven dependency tree That displays both direct and transitive dependencies. 
The SBOM for our solution is only concerned with the direct dependencies as listed in Table~\ref{tab:table-3}.

\begin{table}
\begin{tabular}{ |l| } 
 \hline
 \textbf{Direct Dependencies} \\
 \hline
 javax.servlet:javax.servlet-api:jar:3.1.0 \\
 org.json:json:jar:20160807 \\
 org.apache.maven:maven-model:jar:3.1.0 \\
 com.acme.deveops.ci.common:ci-common:jar:1.0-SNAPSHOT \\
 io.jsonwebtoken:jjwt-api:jar:0.11.2 \\
 io.jsonwebtoken:jjwt--impl:jar:0.11.2 \\
 io.jsonwebtoken:jjwt-jackson:jar:0.11.2 \\
 junit:junit:jar:4.12 \\
 \hline
\end{tabular}
\caption{\label{tab:table-3}Direct Dependencies}
\end{table}

Note that the SBOM is also designed to track internally-developed libraries which are just as important and subject to the same risks as an open source library. The \textbf{ci-common} dependency above is an example of an internal dependency.

\subsubsection{Dependency Reference Database (DRD) Schema}
\label{subsubsec:DRD-schema}
As described above, the SBOM represents an inventory of direct dependencies that are in an application build configuration. 
The dependency reference database (DRD) is central to tracking and controlling the use of open source and internal dependencies. 
At a minimum the dependency reference database will consist of the following tables (see Figure \ref{figure:dependency-database}):

\begin{figure}
    \centering
    \includegraphics[width=\linewidth,trim=0.2cm 16.5cm 6cm 1.5cm, clip]{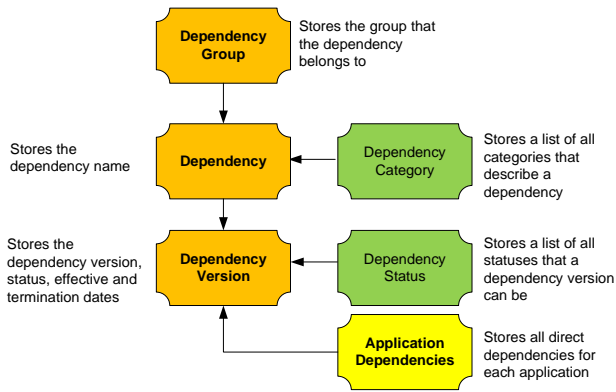}
    \caption{Dependency Reference Database (DRD) entity diagram}
    \label{figure:dependency-database}
\end{figure}

\begin{enumerate}
    \item \textit{Dependency Group} table stores the group or organization name that created the dependency (library). Maven build dependencies for example contain a group such as org.apache, etc while NPM build configurations generally do not so a default value can be substituted for all NPM dependencies.
    
    \item \textit{Dependency Category} table stores the category of a dependency. Dependencies can be categorized by functionality or technology (JSON parser, File Utilities, MVC framework). Categorizing dependencies can help users search and identify dependencies that can be used for specific purposes.

    \item \textit{Dependency} table stores the name of the dependency and the category that the dependency belongs to. (e.g. JSON Parsers, XML Parsers, Date formatters, etc.).

    \item \textit{Dependency Version} table stores the version of the dependency, date that the dependency was introduced into the ecosystem, status of the particular version of the dependency, effective and end date to indicate the usage time frame of the dependency version and the justification for why the dependency was rejected. 

    \item \textit{Dependency Status} table stores the status of the dependency and version. The status is used to determine if the dependency and version is allowable into a build system. Initially there are 4 statuses: Not Reviewed, Approved, Deprecated and Rejected. 

    \item \textit{Application Dependency} table stores which applications use the dependency.
\end{enumerate}

\subsection{Dependency Reference Service (DRS)}
\label{subsec:dependency-reference-service}
\begin{figure} 
    \centering
    \includegraphics[width=\linewidth,trim=2cm 16.5cm 6cm .2cm, clip]{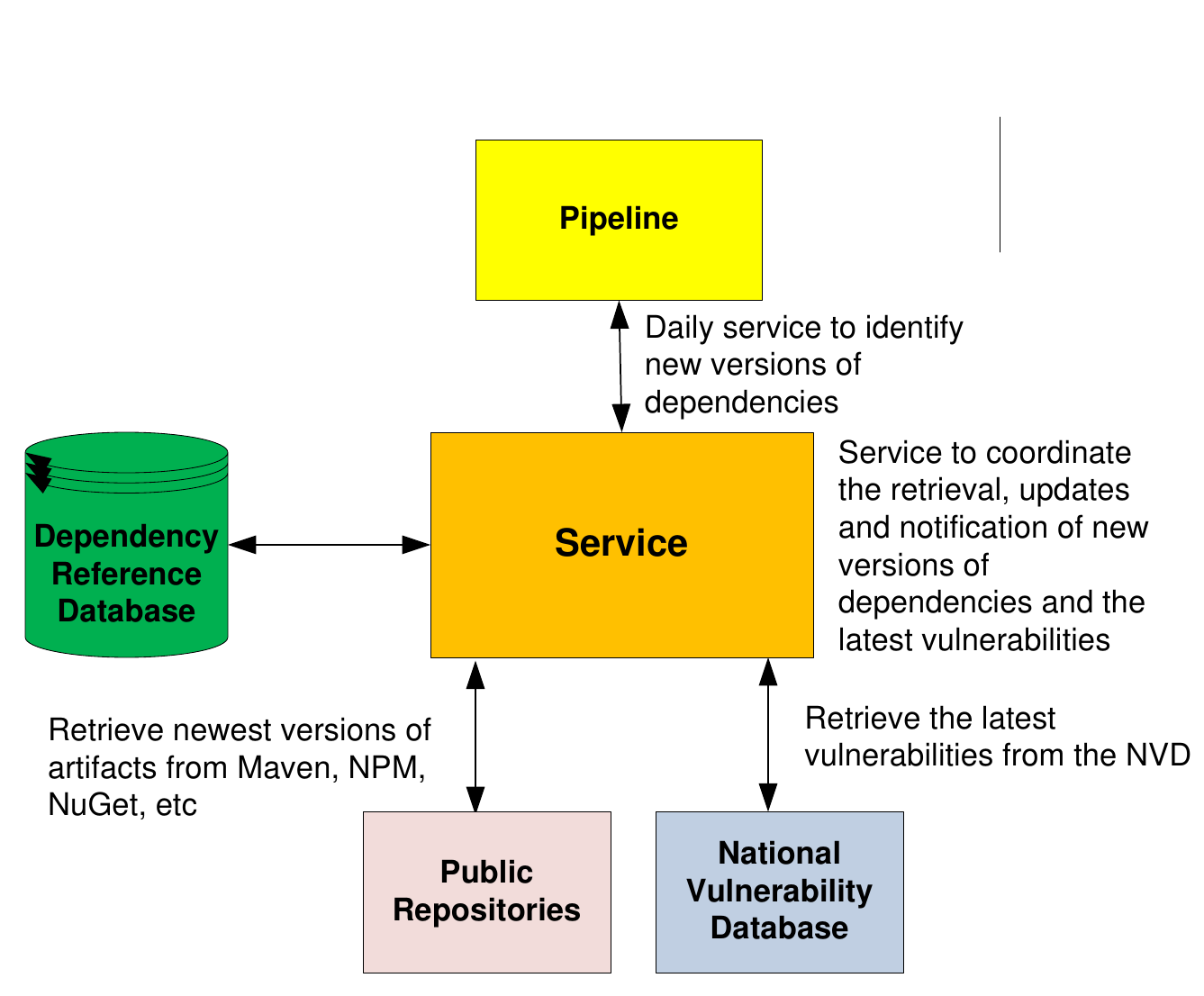}
    \caption{Process of identifying new dependency versions and vulnerabilities}
    \label{figure:identify-new-dependency-versions}
\end{figure}

Figure \ref{figure:identify-dependencies} showed a web service that is triggered by commits in the source control systems to analyze the application against a reference database of dependencies (DRD), with potential documented vulnerabilities. This raises the question: how is the DRD populated? In fact, the DRD is populated by two \textit{services}:
\begin{enumerate}
    \item the web service described in Section \ref{subsec:from-devops-to-devsecops} (Figure \ref{figure:identify-dependencies}), which acts \textit{synchronously} to build requests/commits to source code repositories, and
    \item an \textit{asynchronous} service, whose job is to fill out the vulnerability information (Fig . \ref{figure:identify-new-dependency-versions}).
\end{enumerate}
We explain the interplay in more detail below.

At time $t=0$, when the CI pipeline of Figure \ref{figure:pipeline-we-service-automation}, the DRD is \textit{empty}. As new builds are requested and SBOMs are created (Section \ref{subsubsec:SBOM}), we encounter dependencies for the first time--e.g., a specific version of the \texttt{xerces} XML parser--that were not known to the DRD. Those are thus entered in the DRD with status \texttt{NotReviewed} (see \ref{subsubsec:DRD-schema}); such dependencies may be given a reprieve of a predetermined time duration until their vulnerability status is determined. In parallel, our DRS service can regularly (daily, hourly, etc.) check external sources for security information about the newly added dependency. These include the \textit{National Vulnerability Database}, as well as public repositories that publish vulnerability information in consumable format (see Figure \ref{figure:identify-new-dependency-versions}). In case a query to the NVD, say, identified a vulnerability, the status of the corresponding dependency is updated in the DRD. Note that this does not only concern \texttt{NotReviewed} dependencies: new vulnerabilities are found regularly in libraries that were thought to be safe.

Accordingly, the DRS is invoked periodically to:
\begin{enumerate}
    \item Pull the dependencies from the Dependency Reference Database 
    \item Search the public repositories (Maven, NPM, Nuget, etc.) for the latest versions
    \item Search the National Vulnerability Database (NVD and other sources) for the latest vulnerabilities
    \item Update the Dependency Reference Database with the latest information and notify development teams
\end{enumerate}
Upon notification of the updates the development teams can take the necessary steps to assess and schedule the updates into their applications.

%% file: sections/4-results.tex
\section{Preliminary Results}
\label{sec:results}

In this section, we report on the preliminary results of applying the initial steps of our approach. 
We focus on discussing 1) the current state-of-affairs of open source code usage in the company applications' ecosystem, and 2) we describe two use-cases where better open source control and governance has the potential to save costs in security mitigation strategies and dependency management.    

The organization where our approach is been implemented contains a total of 780 software repositories. 
After the SBOM was populated and analyzed, the SBOM helped identify approximately 1,000 reported vulnerabilities on the 450 versions of dependencies used in the 780 repositories. 
Furthermore, an average of 3 to 5 new vulnerabilities on those versions are being reported per day, indicating a trend of increase overhead in dependency management.

Among the 780 repositories, the portfolio of the companies project is divided into 527 Java repositories, 211 .NET repositories, and 42 JavaScrip repositories. 
Taking only the 527 Java ecosystem into account, once SBOM was put in place, we identified 1,986 different Java open source libraries versions, including solely direct dependencies. 
Table~\ref{tab:breakdown-java-libraries} shows the breakdown of libraries for the 15 most common library domains in the company application ecosystem.
One can observe a plethora of options currently maintained by the company developers. 
For instance, the SBOM showed that across all Java projects, developers maintain 60 different Web frameworks, 54 different logging libraries, and 52 different Database connectivity libraries. 
This shows that open source usage, if left unchecked, can lead multiple teams of developers maintaining dozens of different libraries for the same purpose. 
Each library incurs in its own maintenance costs: each library contains its own API, are maintained by different teams with potentially different update policies, and are subject to their own set of reported vulnerabilities.

\begin{table}[]
    \centering
    \caption{Breakdown of Java libraries used in the company application ecosystem per library domain}
    \label{tab:breakdown-java-libraries}
    \begin{tabular}{l r}
        \toprule
        \textbf{Library Domain} & \textbf{\# of Different Libraries}  \\
        \midrule
        Web Frameworks & 60 \\
        Logging  & 54 \\
        Database Connectivity  & 52 \\
        REST framework  & 43 \\
        SOAP  & 30 \\
        PDF  & 34 \\
        Email  & 21 \\
        ORM  & 21 \\
        XML Parser  & 18 \\
        Encryption & 16 \\
        JSON Parser & 12 \\
        Date/Time Parser & 8 \\
        Charting & 7 \\
        Caching & 6 \\
         \bottomrule
    \end{tabular}
\end{table}

We present in Tables~\ref{tab:xml-dependencies} and~\ref{tab:json-dependencies}, a list of currently used Java libraries for XML and JSON parsers per version, and their respective number of Vulnerabilities, as reported by the company security tooling.  
We highlight the dependencies that contain at least one vulnerability in bold. 
The highlighted XML dependencies account for a total of 20 vulnerabilities affecting a total of 29 library versions. 
Notice that the remaining XML parsers, used by the company applications, contain no vulnerabilities. This outlines two problems: the use of multiple dependencies that provide the same basic functionality and several of them are vulnerable. 
It would make sense then that by reducing the number of XML dependencies to possibly one or two that are not vulnerable then that will eliminate the vulnerabilities entirely. It would also reduce maintenance and training costs since developers would not be required to know the intricacies of every XML parser.

We found a similar picture when analyzing JSON parsers (see Table~\ref{tab:json-dependencies}). 
Two libraries reported 3 vulnerabilities across 11 different versions, however, all other libraries used by the company application ecosystem reported zero vulnerabilities, at the time of analysis. 
There were many other similar occurrences across that organization's ecosystem. Without an SBOM and automated pipelines to track the dependencies being introduced there would be no easy way to identify and control open source usage. 

\begin{table}
\caption{Java XML parser dependencies found in the application ecosystem of studied company. The column ``\# of Vulns'' showcases the total number of vulnerabilities affecting all dependency versions.}
\label{tab:xml-dependencies}
\input{tables/xml-json-dependencies}

\end{table}

\begin{table}
\caption{Java JSON parser dependencies found in the application ecosystem of studied company. The column ``\# of Vulns'' showcases the total number of vulnerabilities affecting all dependency versions.}
\label{tab:json-dependencies}
\input{tables/json-dependencies}

\end{table}

%% file: tables/xml-json-dependencies.tex
\begin{tabular}{l r r r} 
\toprule
\textbf{XML and JSON Parsers} & \# \textbf{Vulns} & \# \textbf{Versions} \\% & \# \textbf{Repos} \\
\midrule
\textbf{xstream} & 6 & 4 \\%& 6 \\
\textbf{xmlsec} & 6 & 3 \\%& 3 \\
\textbf{jackson-dataformat-xml} & 3 & 13 \\%& 32 \\
\textbf{dom4j} & 1 & 2 \\%& 1 \\
\textbf{jdom} & 1 & 1 \\%& 1 \\
\textbf{xom} & 1 & 1 \\%& 1 \\
\textbf{xmlbeans} & 1 & 3 \\%& 2 \\
\textbf{xalan} & 1 & 2 \\%& 1 \\
xmlschema & 0 & 1 \\%& - \\
xerces & 0 & 1 \\%& - \\
sax & 0 & 1 \\%& - \\
xml-aps & 0 & 2 \\%& - \\
xmlpublic & 0 & 1 \\%& - \\
aalto-xml & 0 & 1 \\%& - \\
javax.xml.stream & 0 & 1 \\%& - \\
xmlpull & 0 & 1 \\%& - \\
xpp3_min & 0 & 1 \\%& - \\
xmlsec & 0 & 1 \\%& - \\
\bottomrule
\end{tabular}

%% file: tables/json-dependencies.tex
\begin{tabular}{l r r} 
\toprule
\textbf{JSON Parsers} & \# \textbf{Vulns} & \# \textbf{Versions} \\% & \# \textbf{Repos} \\
\midrule
\textbf{json-smart} & 2 & 4 \\ %
\textbf{gson} & 1 & 7 \\%& 22 \\
json & 0 & 10 \\%& - \\
json-lib & 0 & 3 \\%& - \\
json-simple & 0 & 2 \\%& - \\
json-path & 0 & 1 \\%& - \\
javax.json-api & 0 & 1 \\%& - \\
tapestry-json & 0 & 1 \\%& - \\
wink-json4j & 0 & 1 \\%& - \\
jakarta.json & 0 & 1 \\%& - \\
json4s-core & 0 & 1 \\%& - \\
jsonschema2pojo-core & 0 & 1 \\%& - \\
\bottomrule
\end{tabular}

%% file: sections/5-discussion.tex
\section{Discussion}
\label{sec:discussion}
\subsection{Where to from here?}
\label{subsec:next-steps}
The envisioned DevSecOps pipeline described in Section \ref{subsec:from-devops-to-devsecops} is partially implemented. The two services described in Figures \ref{figure:identify-dependencies} and \ref{figure:identify-new-dependency-versions} have been implemented and are fully operational, providing us with the results described in Section \ref{sec:results}. However, the \textit{policies} that we mentioned in sections \ref{subsec:from-devops-to-devsecops} and \ref{subsec:dependency-reference-service} have not yet been implemented. Example policies include deciding what to do when an application is found that depends on a library with known vulnerabilities. Alternatives include:
\begin{enumerate}
    \item Informing a designated representative of the team that committed the code triggering the dependency check of the security vulnerabilities, without failing the pipeline.
    \item Informing the designated representative of the vulnerabilities, giving the build a reprieve to migrate to a different library or to a patched library version, if one exists, possibly blocking the OPS part of DevSecOps in the meantime.
    \item Informing the designated representative of the vulnerabilities, and failing the build right away.
\end{enumerate}
A similar range of alternatives may apply to dependencies encountered the first time (status \texttt{NotVerified}). For the time being, we have not implemented such policies/rules, for several reasons:
\begin{enumerate}
    \item We have to \textit{design} such policies, in a consensual way that balances safety and agility,
    \item We have to \textit{educate} developers about our DevSecOps approach, and the rationale behind any policies we might come up with,
    \item The DRD has to have 'enough coverage' for the policies to make sense,
    \item We have to set-up the appropriate processes and structures to make this work.
\end{enumerate}
In the remainder of this section, we will touch upon some of these issues.

\subsection{Developer Education}
\label{subsec:developer-education}
Needless to say, security is everybody's business and developers are prime stakeholders in the process. While we strive to minimize the disruption to their work flow, having one's build fail, \textit{after the fact}, because they used the wrong library, is wasteful and can be frustrating to developers. Thus, it is important that developers be: 1) educated about the rationale between the process, and 2) perhaps trained, to the extent that it is possible, to avoid having their builds fail.

Thus, a series of training workshops need to be created to provide the necessary education and awareness to the development teams. Some of the topics to be covered included:

\begin{itemize}
    \item An introduction of the new policies and the reason why the organization has undertaken this effort. The developers need to understand the importance of controlling open source software usage and why it will benefit the organization as a whole.
    \item A set of guidelines for development teams to follow and ensure that all open source software that is introduced into the application ecosystem is done in a responsible, systematic and governed way.
    \item How to use the dependency reference web application and to always consult the web application to find the software and versions that are sanctioned for use prior to looking outside.
    \item Developers need to be taught to focus more attention to what software they are using and to ensure that their build configurations are always kept up to date. 
\end{itemize}

\subsection{A Software Vetting Process}
\label{subsec:new-software-vetting-process}
The Dependency Reference Service (DRS) of Figure~\ref{figure:identify-new-dependency-versions} was shown to pull its information from the \textit{National Vulnerability Database} (NVD) as well as public repositories where OSS is maintained. 
However, it is unrealistic to assume that \textit{all} the quality-related information about OSS can be found: 1) in public repositories, and 2) in query-able fashion. First, the NVD cannot possibly be exhaustive. Second, security vulnerabilities are but one reason, among many, for not using an OSS library. Finally, we will likely \textit{not} find direct quality metrics, although machine learning techniques may be used to correlate quality with available metrics.
For these reasons, and for the short/mid-term, the Dependency Reference Database (DRD) will be populated partially \textit{manually}, by a \textit{governance committee} whose job is to vet new dependencies or dependency versions that are encountered for the first time in a build, and that may not have been found in the NVD. Such a committee, which should include subject matter experts and security specialists, will need to ask the following questions about any new OSS component detected by the build dependency service (Fig. \ref{figure:identify-dependencies}):
\begin{enumerate}
    \item \textit{Is there an approved component that provides the same functionality?} One of the key functionalities of this effort is to ensure that software with duplicate capabilities are kept out of the ecosystem or at least kept to a minimum. 

    \item \textit{How many vulnerabilities and what is the severity?} Another key functionality of this effort is to ensure that any new software that is introduced into the ecosystem contains no critical vulnerabilities. Source composition analysis tools, if available, can easily provide that information otherwise other sources such as the National Vulnerability Database can be consulted.

    \item \textit{Is it actively maintained?} There are many open source projects that are not actively maintained resulting in outdated software that in most cases can lead to unpatched vulnerabilities and other quality issues. A thorough research on the commit history of that software, if available, should be performed. 

    \item \textit{Who is the maintainer?} If the software is being maintained by well-known organizations such as Google, The Apache Foundation and IBM then there can be some assurance that the development team has resources to respond to reports of vulnerabilities, write extensive test suites, and respond to the community feedback, than repositories maintained by single developers.  

    \item \textit{How do they view security?} Viewing the commit logs and bug reports for the software, if they exist, will shed light on how the maintainer views security. Quick remediation strategies and well-documented updated patches are a good indicator that security and quality are important to the maintainer.

    \item \textit{What is their issue history?} Viewing the bug logs, if they exist, is a good indicator of the quality history of the software. 
\end{enumerate}

As new software versions are approved, older versions of an OSS component may be put into a deprecated status. For high security threat libraries, the deprecation process will happen in a rapid, blacklisting effort blocking further use sooner rather than later as determined by management. For lower threat libraries, the phase-out process will happen within a carefully orchestrated process to support those apps that need the library.

\subsection{Research, Categorize, and Reduce}
\label{subsec:research-categorize-reduce}
The preliminary results shown in Section \ref{sec:results} showed that we have \textit{way too many} OSS libraries in our application portfolio the perform similar or identical functions, security vulnerabilities notwithstanding. Recall from table \ref{tab:breakdown-java-libraries} the number of web frameworks (60!), logging frameworks (54), or database connectivity frameworks (52) in use at the company. The multiplicity of libraries reduces knowledge sharing between teams, and increases maintenance costs.

Thus, one of our goals is to research each dependency and version that is found in our applications ecosystem, and reduce the number to smaller (sub)set of approved components. This will be part of the mandate of the governance committee mentioned above. Some of the criteria that should be considered are:

\begin{enumerate}
    \item \textit{Identify and eliminate duplicate functionality:} Identify the software that provide the same functionality. For instance, there may be two or three model, view, controller (MVB) frameworks that are currently found to be used in the applications. The question should be asked: "Do we really need multiple MVC frameworks in those software projects?". After some amount of research has been performed the answer may be "No" therefore one (or few) MVC framework should be decided based on factors such as capabilities, best fit, number of vulnerabilities, etc. 

    \item \textit {Preference of native language features versus open source libraries:} In some cases, the best architecture practice may be to remove an open source or home-grown library altogether in favor of the native functionality of a given software development language. A common evolution in software language development has seen elements incorporated into the native language that may have been addressed in an open source component.  There are many instances where certain capabilities in a language were either not included or was not performant enough which resulted in it's inclusion into open source. If the functionality is the same then it is preferable to use the native language rather than the open source thus eliminating a possible open source risk. 

    \item \textit{Choose the open source dependencies that are least vulnerable.} Source Composition Analysis (SCA) tools can help identify the vulnerabilities and their respective severity levels which will aid in choosing the best software to use.
\end{enumerate}

Note that the SBOM in this solution not only contains open source dependencies it also tracks internally-built dependencies as well. Initially it might prove easier to identify and review the internal dependencies as there may be multiple versions of these dependencies in the ecosystem. Since this software is usually built in-house it is much easier to consolidate and standardize on one or two versions of these dependencies before reviewing the open source which may be more challenging.

%% file: sections/6-relatedwork.tex
\section{Related Work}
\label{sec:related-work}

In this section, we discuss the related literature from two main perspectives. 
First, we discuss works that have focused in describing the challenges in open-source software usage. 
Then, we discuss industrial reports that propose approaches for mitigating OSS usage problems.

\subsection{Challenges of Open Source Usage}
\label{sub:open-source-challenges}

A wide range of works have focused in challenges of open-source dependency management, from selecting well-maintained software libraries~\cite{Abdalkareem:17:Trivial,Costa:21:Breaking,Mujahid:21:Centrality,Coelho:20:Maintained}, to establishing good update policies~\cite{Jafari:21:DependencySmells,Cogo:19:Downgrades,Bogart:16:Breaking}.
Abdalkareem et al. investigated why developers tend to select packages even to implement trivial programming tasks~\cite{Abdalkareem:17:Trivial}.
According to the authors, developers select trivial packages because they perceive them to be well implemented and tested, when in fact 45\% of the packages did not even have explicit tests. 
As developers tend to select libraries based on their popularity in the community, Mujahid et al. discuss the problems of current popularity metrics~\cite{Mujahid:21:Centrality}. 
In their work, they find that major popularity metrics (dependencies, downloads) are not well-correlated to what the community actually uses, which could lead developers to selecting outdated legacy packages for their projects.
Similarly, Coelho et al. proposed a data-driven approach to identify well-maintained projects in Github~\cite{Coelho:20:Maintained}, showing that 16\% of studied projects tend to be unmaintained for more than a year. 

To effectively use open-source code, developers have to constantly update their dependencies to get the latest big fixes and vulnerability patches. 
Kula et al. report that developers, in fact, do not frequently update their dependencies, with 81\% of studied projects relying on outdated library versions~\cite{kula:18:Update}. 
On a similar note, Jafari et al. discuss challenges of dependency management  ~\cite{Jafari:21:DependencySmells}. 
They identify seven recurring dependency management issues, named dependency smells, which can on the long term increase the number of bugs and vulnerabilities of a package. 
The authors quantified the occurrence of such dependency smells and reported that the vast majority of libraries (92\%) has at least one dependency smell in their manifest files.  
Côgo et al.'s work on dependency downgrades also provides great insights on the challenges of dependency management and shows that dependency downgrades can be a workaround for update-related issues in client code \cite{Cogo:19:Downgrades}. 
This is partially explained by Bogart et al.'s survey with developers from 18 ecosystems which shows that updates in packages often break the code of their client, requiring modifications in the client code \cite{Bogart:16:Breaking}.

Many researchers have studied the presence of software vulnerabilities in open source dependencies \cite{Decan:18:NPM,decan:19:networkevolution,kikas:17,Alfadel:21:Python,Imtiaz:21:Reporting}. 
Decan et al. investigated the life-cycle of vulnerabilities in npm, showing that the number vulnerabilities are growing over time and take in median 24 months to be discovered~\cite{Decan:18:NPM}.
Alfadel et al. replicated their study in the Python ecosystem and reported that Python vulnerabilities take even longer to be found (3 years in median), and are frequently made public before there is no remediation strategy in place, increasing the risk of exploitation~\cite{Alfadel:21:Python}.
To make matters worse, Kula et al~\cite{kula:18:Update} also report that developers are frequently unaware that their projects depend on vulnerable dependencies. 
In their survey, 69\% of the respondents were unaware their project depended on vulnerable library. 
To account for this widespread problem of software vulnerabilities, the open source community has also proposed a few solutions~\cite{Alfadel:21:Dependabot}. 
Approaches such as Dependabot~\cite{Alfadel:21:Dependabot} have been well-received by the community as they help developers keep their dependencies up-to-date and raise awareness for the occurrence of software vulnerabilities.

All the above mentioned works have been conducted primarily in the open source world. 
The authors focused on studying the challenges of reusing open source code in open source projects. 
Our work complements the above mentioned literature by bringing an industrial perspective to the problem. 
We discuss a framework that helps mitigate the risks of using open source code in a large company, and each component of our DevOps pipeline can also be employed by teams working in open source projects.

\subsection{Industrial Approaches to Mitigate the Risks of Using Open Source Code}
\label{sub:industrial-reports}

While the interest on the challenges of open source usage in industry has spiked in the last several months with the recent cases of log4shell~\cite{Log4Shel70:online}, we found that industrial reports on OSS control and governance is still largely unexplored. 
Still, there are some recent works that focuses on the challenges of using open source in an industrial scale, which we describe next.   

Plate et al. investigated how application vendors in industry determine if a reported vulnerability puts their application at risk~\cite{Plate:15:ImpactAssessment}. According to the authors, current decision making process relies mostly on vulnerability descriptions and expert knowledge, hence, being time-consuming and error-prone. 
Dann et al. evaluated the robustness of popular code vulnerability scanners to changes in code packaging strategies, commonly employed in industry~\cite{Dann:21:Scanners}. The study also showed that major technology companies, such as SAP, rely extensively on open source code, with 86\% of dependencies of proprietary projects coming from open source projects. 
Ponta et al discussed approaches for dealing with bloated dependencies~\cite{Ponta:21:Bloated}. 
They evaluated three approaches that use code analysis to identify redundant code (DepClean, Maven Shade, and Pro Guard) and were all similarly effective in de-bloating applications. The authors also report that such de-bloating strategies were effective in reducing the attack surface of two industrial applications.
Pashchenko et al. discussed the overhead of reported dependencies with un-exploitable vulnerabilities in \cite{Pashchenko:20:Vuln4Real}. They also survey developers on dependency management in \cite{Pashchenko:20:SecurityManagement} and find that most developers find that software composition analysis (SCA) tools generate many irrelevant and low-priority alerts. 

Our study continues the thread of above mentioned works in improving how open source code is used in an industrial environment. 
While still in its early phase, the proposed approach attempts to cover multiple aspects of open source control and governance, and can be employed on different industrial settings to mitigate the risks that come with open source usage.

%% file: sections/7-conclusion.tex
\section{Conclusion}
\label{sec:conclusion}

In this paper, we discussed the risks related to using open source libraries and proposed an approach to help mitigate such risks, with better control of OSS code reuse. 
Our approach focuses on using automation to reduce open source risks, by including a control mechanism in the DevOps pipeline to help document all the dependencies used in a company's software ecosystem and vet libraries that are deemed risky by experts and stakeholders.  
This process is currently being implemented by a large company with more than 400 developers and we discuss the potential for saving costs by reducing library redundancy and dependency overhead. 
The implementation of our approach, however, is still ongoing.  
There are challenges pertaining to how to communicate this new process to developers to increase awareness of the risks caused by the lack of open source governance. 
In our future work, we plan to report on a quantitative evaluation of the approach's potential for mitigating dependency management and its impact on cost savings; we will also report on the qualitative feedback from software developers.